\documentclass{article}

\usepackage{PRIMEarxiv}

\usepackage[utf8]{inputenc} 
\usepackage[T1]{fontenc}    
\usepackage{hyperref}       
\usepackage{url}            
\usepackage{booktabs}       
\usepackage{amsfonts}       
\usepackage{nicefrac}       
\usepackage{microtype}      
\usepackage{lipsum}
\usepackage{fancyhdr}       
\usepackage{graphicx}       
\graphicspath{{media/}}     
\usepackage{amsmath}

\title{Understanding planetary context to enable life detection on exoplanets and test the Copernican principle
\thanks{\textit{\underline{Citation}}: 
\text{Krissansen-Totton, J., et al. (2022). Nature Astrononmy, 6, 189-198. https://doi.org/10.1038/s41550-021-01579-7}} 
}

\author{
  Joshua Krissansen-Totton, Maggie Thompson, Max L. Galloway, and Jonathan J. Fortney  \\
  Department of Astronomy and Astrophysics, CA, USA \\
  University of California, Santa Cruz \\
  Santa Cruz, CA, USA\\
  \texttt{jkt@ucsc.edu} \\
}

\begin{document}
\maketitle

\begin{abstract}
The search for life on exoplanets is motivated by the universal ways in which life could modify its planetary environment. Atmospheric gases such as oxygen and methane are promising candidates for such environmental modification due to the evolutionary benefits their production would confer. However, confirming that these gases are produced by life, rather than by geochemical or astrophysical processes, will require a thorough understanding of planetary context, including the expected counterfactual atmospheric evolution for lifeless planets. Here, we evaluate current understanding of planetary context for several candidate biosignatures and their upcoming observability. We review the contextual framework for oxygen and describe how conjectured abiotic oxygen scenarios may be testable. In contrast to oxygen, current understanding of how planetary context controls non-biological methane (CH$_4$) production is limited, even though CH$_4$ biosignatures in anoxic atmospheres may be readily detectable with the James Webb Space Telescope. We assess environmental context for CH$_4$ biosignatures and conclude that abundant atmospheric CH$_4$ coexisting with CO$_2$, and CO:CH$_4$ << 1 is suggestive of biological production, although precise thresholds are dependent on stellar context and sparsely characterized abiotic CH$_4$ scenarios. A planetary context framework is also considered for alternative or agnostic biosignatures. Whatever the distribution of life in the Universe, observations of terrestrial exoplanets in coming decades will provide a quantitative understanding of the atmospheric evolution of lifeless worlds. This knowledge will inform future instrument requirements to either corroborate the presence of life elsewhere or confirm its apparent absence.
\end{abstract}

\section{Main}
In the coming decades, observational astronomy will be uniquely positioned to establish whether life is common in the Universe, or whether Earth’s rich and diverse biosphere is a cosmic aberration. This final test of the Copernican Principle will begin with new instruments such as the James Webb Space Telescope (JWST) and the ground-based Extremely Large Telescopes of the 2020s. However, surveying a statistically meaningful sample of planets for signs of life will require a direct-imaging flagship mission, such as the $\sim$6m infrared/optical/ultraviolet (IR/O/UV) space telescope recommended by the Astrophysics 2020 Decadal Survey$^1$, or the in-development LIFE interferometer telescope concept$^2$.

Knowing what to look for is an ostensibly immense challenge to exoplanet life detection given the limited knowledge (sample size N=1) of the possible diversity of biochemistries that may exist in nature. However, attempts to generalize our understanding of life can help guide the search$^3$. Schrödinger$^4$ provided a working definition of life suitable for framing the search for biosignatures beyond Earth: life is a non-equilibrium system that feeds upon free energy and contains the instructions for its own self-replication, in what Schrödinger termed a molecular ‘codescript’. This definition was, in effect, a prediction that successfully anticipated and inspired the subsequent discovery of the structure of DNA, the codescript for Earth life$^5$. The ability for non-equilibrium systems to specify their own self-replication is, in turn, a prerequisite for open-ended Darwinian evolution$^{6,7}$.

If life exists elsewhere in the Solar System, then it may be possible to uncover direct evidence of the universal thermodynamic structures of life envisaged by Schrödinger. Such evidence could, in principle, be independent of the broader planetary context within which it was found. For example, motile structures in the plume material of Enceladus would provide proof of a Darwinian process, or a sample returned from Mars full of hopanes or complex, chiral organics would imply molecular specificity so thermodynamically improbable that it could only be a product of out-of-equilibrium living structures.

In contrast, for exoplanets, astronomers are unlikely to uncover direct evidence of life’s thermodynamic structures—only the planetary context will be accessible to future instruments. The challenge of extrasolar astrobiology is thus to identify biospheres through the expected ways in which self-reproducing, out-of-equilibrium systems modify their planetary environments. Understanding of the thermodynamic requirements for life—specifically the exploitation of free energy—can be leveraged to focus attention on the metabolic waste products that would (1) be plausibly produced by a productive surface biosphere and (2) subsequently accumulate to detectable levels. Understanding of astrophysics, geophysics and geochemistry must then be invoked to rule out non-biological sources for such waste products. In this way, the search for life is inseparable from fully characterizing a planetary environment; that is, to both look for the by-products of metabolism and to rule out the non-biological scenarios that might otherwise explain observations.

Here we first illustrate a contextual approach to exoplanet biosignature detection using oxygen as the best-understood example, and argue that oxygen remains a compelling biosignature in some planetary contexts. However, methane (CH$_4$) biosignatures will be more easily detectable in roughly the next five years and are comparatively understudied. Thus, we make a case for developing a contextual framework for interpreting CH$_4$ detections in terrestrial planet atmospheres to determine how to distinguish biological and abiotic CH$_4$. Finally, we argue that comparable context-dependent approaches could be applied to other biosignatures. Our focus is on atmospheric biosignatures, given their imminent observability, but a similar contextual approach may be applied to surface or temporal biosignatures$^{8,9}$.

\section{Oxygen as a case study in contextual arguments for life}
Molecular oxygen (O$_2$), or its photochemical by-product ozone (O$_3$), was considered a promising potential biosignature long before the discovery of exoplanets$^{10,11}$. Arguments for oxygen as a biosignature can be boiled down to two premises. First, regardless of specific biochemistry, any surface life that develops the capacity for oxygenic photosynthesis will possess an immense evolutionary advantage over other energy metabolisms. Chemotrophic life depends on geochemical sources of free energy and anoxygenic photosynthesis is similarly limited by the availability of electron donors such as H$_2$ or Fe$^{2+}$. In contrast, oxygenic photosynthesis requires only water and carbon dioxide (CO$_2$), substrates that are virtually unlimited on planets with habitable surfaces:
\begin{equation}
\text{CO$_{2}$ + H$_{2}$O $\rightarrow$ CH$_{2}$O + O$_{2}$}
\end{equation}

Second, non-biological production of large amounts of oxygen is unlikely. This second premise has received considerable attention in the astrobiology literature, to the point where there are now detailed reviews assessing scenarios for non-biological oxygen production, their plausibility and possible contextual clues to establish biogenicity (or lack thereof)$^{12}$. While understanding of atmospheric oxygenation remains incomplete, oxygen is the best studied example of how to assess biosignatures through planetary context.

Much of the focus has so far been on so-called photochemical oxygen false positives. The photodissociation of oxygen-bearing molecules such as CO$_2$ can, in principle, result in steady-state atmospheres rich in O$_2$ or O$_3$ (refs. $^{13,14,15}$). The viability of photochemical oxygen production depends on the spectral energy distribution of the host star, as well as on catalytic recombination processes in both the atmosphere and on the surface$^{13,14,15,16,17,18,19}$. Such photochemically produced O$_2$ may be diagnosable from the presence or absence of other atmospheric constituents (for example CO), as well as from the stellar context$^{12}$.

The ways in which broader planetary context controls the long-term build-up of oxygen due to slight imbalances in sources and sinks is less understood. Oxygen build-up on planets around low mass stars due to extensive hydrogen escape during the pre-main sequence has been suggested$^{20}$, and coupled models have explored whether such oxygen build-up could overwhelm magma ocean sinks$^{21,22,23}$. The possibility of oxygen accumulation due to water loss in atmospheres with low non-condensable inventories has also been proposed$^{24,25}$. In general, however, understanding of whole-planet redox evolution remains rudimentary and untested.

Krissansen-Totton et al.$^{26}$ modelled the evolution of whole planetary contexts, from magma ocean to temperate geochemical cycling, with the goal of anticipating non-biological oxygen accumulation. The model attempts to include all non-biological processes that add or remove oxygen from the atmosphere. The thermal evolution of the interior is explicitly coupled to surface volatile evolution to self-consistently model outgassing and other crustal–atmosphere exchange processes. While lifeless Earth twins are not predicted to possess oxygen-rich atmospheres in this framework, several scenarios were identified whereby oxygen could build up to detectable levels without life if initial volatile inventories differed dramatically from that of Earth (Fig. 1). Crucially, such whole-planet modelling can help identify the contextual observations required to rule out non-biological oxygen. For example, terrestrial planets with large water inventories (tens to hundreds of Earth oceans) will rapidly cease to produce new crust due to the pressure overburden of the surface ocean$^{27,28}$. This inhibits all oxygen sinks such as magmatic outgassing and serpentinization since these processes are limited by fresh crustal production, and this, in turn, allows for long-term oxygen accumulation via hydrogen loss to space$^{22,26}$. While the precise water inventory threshold for oxygen sink suppression is fuzzy, this conclusion is qualitatively insensitive to crustal composition, mantle redox or tectonic regime$^{26}$. Fortunately, since tens of Earth oceans are required in this scenario, the presence of any subaerial land would preclude the water depths$^{29}$ required for waterworld false positives. Time-resolved photometry could be used to detect such an ocean–land dichotomy to rule out waterworld oxygen false positives$^{30,31,32}$.

\begin{figure}
  \centering
  \includegraphics[width=15cm]{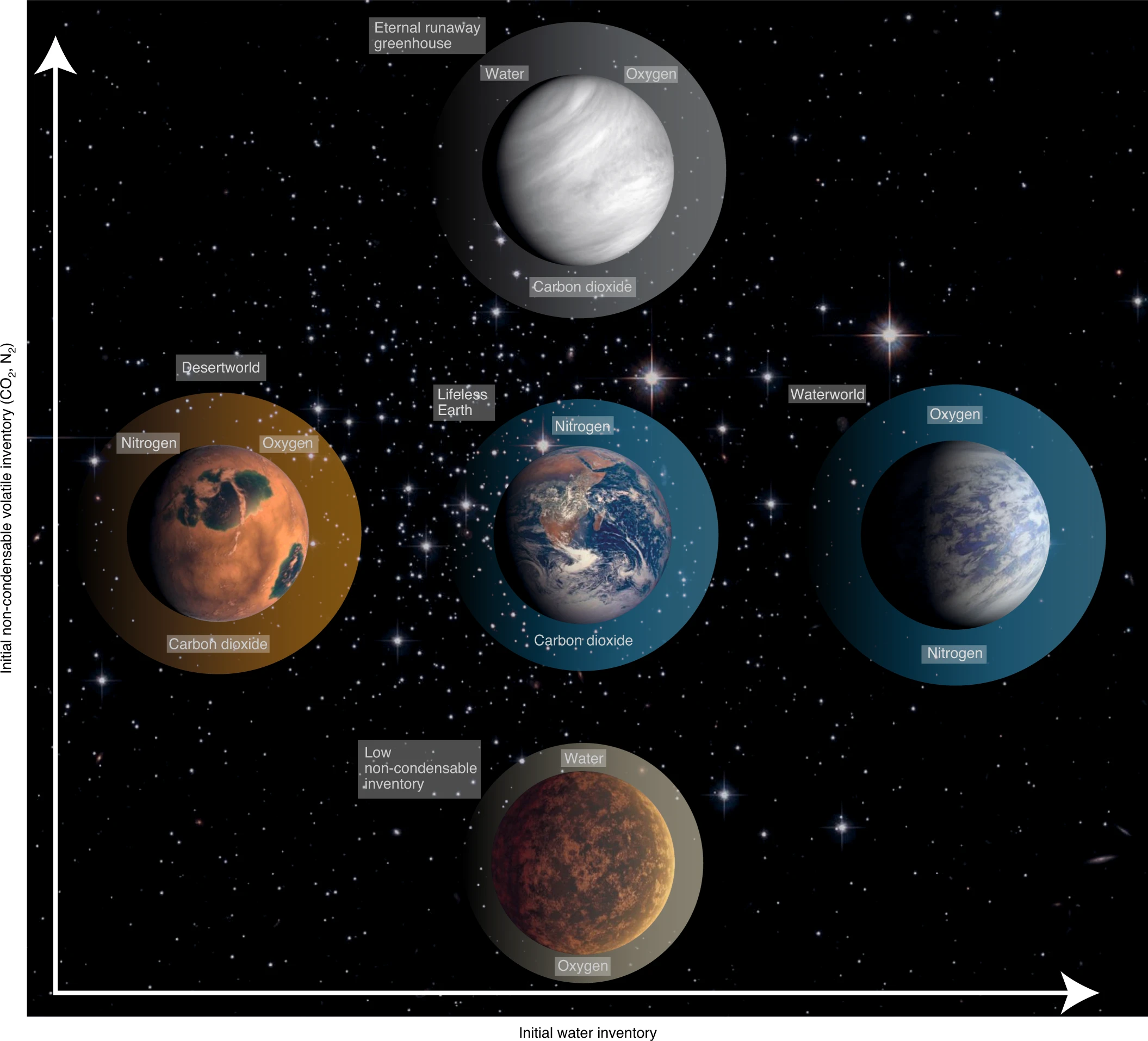}
  \caption{\textbf{Initial volatile inventories may influence the likelihood of non-biological O$_2$ accumulation.} Non-biological O$_2$ build-up is improbable for planets around Sun-like stars with initial water and non-condensable gas (CO$_2$, N$_2$) endowments comparable to those of Earth$^{26}$ (centre). However, large departures from Earth-like inventories can alter planetary redox evolution. For example, oxygen sinks may be suppressed on planets with tens of Earth oceans (waterworlds, right), whereas high CO$_2$/H$_2$O endowments can lead to eternal runaway greenhouse states where oxygen may build up, even within the habitable zone (top). Conversely, low non-condensable inventories can accelerate oxygen build-up via hydrogen escape$^{24}$ (bottom), and desiccated planets may accumulate oxygen early in their evolution$^{26}$ (left). Credit for planet images: NASA/JPL-Caltech.}

\end{figure}

While many simplifying assumptions are necessary, such models can be used to begin mapping out the likelihood of non-biological oxygen production as a function of astrophysical variables, such as planet–star separation, stellar age and initial volatile inventories and so on. (Fig. 2). These statistical relationships could be validated with future instruments.

\begin{figure}
  \centering
  \includegraphics[width=16cm]{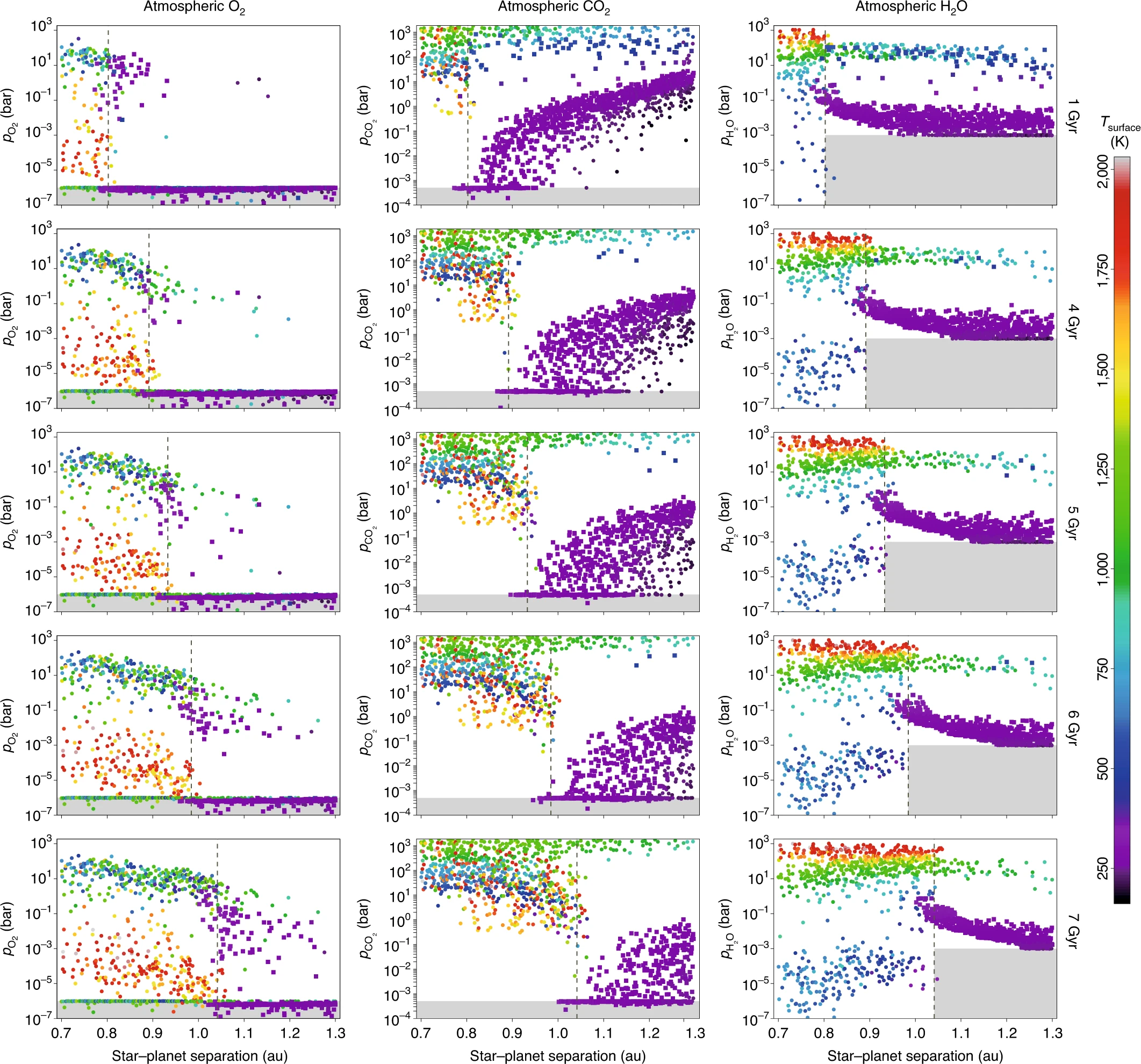}
  \caption{\textbf{Time evolution of atmospheric O$_2$, CO$_2$ and H$_2$O vapour as a function of planet–star separation for a sample of simulated lifeless planets.} The colour scale shows the mean surface temperature, and the black dashed line shows the runaway greenhouse limit for an Earth-like albedo, which evolves with stellar luminosity (a G star is assumed). Squares denote non-zero surface liquid water inventories, whereas circles show model runs with uninhabitable surface conditions. Each row shows the atmospheric composition at a different time in the main-sequence lifetime. The simulated planet population, taken from ref.$^{26}$, has a wide range of initial volatile inventories and parameter values that govern atmosphere–interior exchange of volatiles. Models such as this can be used to predict trends in non-biological oxygen accumulation alongside their contextual clues. The grey shaded regions denote numerical limits; lower abundances may be realized but fluxes cut off here for numerical efficiency.}

\end{figure}

\section{Observational prospects for O$_2$ biosignatures in the 2020s}

Biogenic oxygen will probably not be accessible to JWST for known targets$^{33,34,35,36}$. Although biological oxygen detection might be possible with high-resolution spectrographs from the ground-based Extremely Large Telescopes$^{37,38,39}$, this will require many coadded transits. Nevertheless, JWST will allow coupled models of planetary atmosphere–interior evolution to be tested and validated. Excessive non-biological oxygen build-up on highly irradiated planets (tens to hundreds of bars of O$_2$) could be observable with JWST$^{35,40}$, and so the presence or absence of oxygen on highly irradiated planets may directly constrain escape physics and surface–atmosphere interactions. If the TRAPPIST-1 planets$^{41}$ and other highly irradiated terrestrials have atmospheres, then a many-transit deep dive with JWST$^{42}$ to constrain their compositions would be extremely valuable for interpreting any future oxygen detections on habitable zone planets with next-generation instruments.

A timely return to Venus will also provide critical information for any future exoplanet oxygen detections. Venus’s modern atmosphere contains virtually no O$_2$, despite D/H ratios suggestive of excessive water loss$^{43}$. Understanding Venus’s climate evolution, particularly the fate of water and oxygen, is a prerequisite for interpreting oxygen on potentially habitable exoplanets. Observations of surface mineralogy, trace gas inventories, and tectonic recycling with the upcoming DAVINCI$^{44}$, VERITAS$^{45}$ and EnVision$^{46}$ missions could dramatically improve constraints on crustal oxygen sinks and atmospheric escape physics. Venus is, after all, the only other Earth-sized planet accessible to in situ analysis, and thus presents a unique opportunity to advance understanding of soon-to-be-observed terrestrial exoplanets.

\section{CH$_4$ biosignatures and early Earth analogues}

While oxygenic photosynthesis seems like the logical evolutionary endpoint to oxygen-respiring heterotrophs, there is no guarantee that oxygen-rich atmospheres are ubiquitous, even if life is common. Oxygenic photosynthesis could have been a one-off evolutionary event not frequently repeated, or oxygen-rich atmospheres may not easily accumulate$^{47}$. This motivates the search for alternative biosignatures. In this context, the history of atmospheric modification by life on Earth is a useful guide (Fig. 3).

\begin{figure}
  \centering
  \includegraphics[width=15cm]{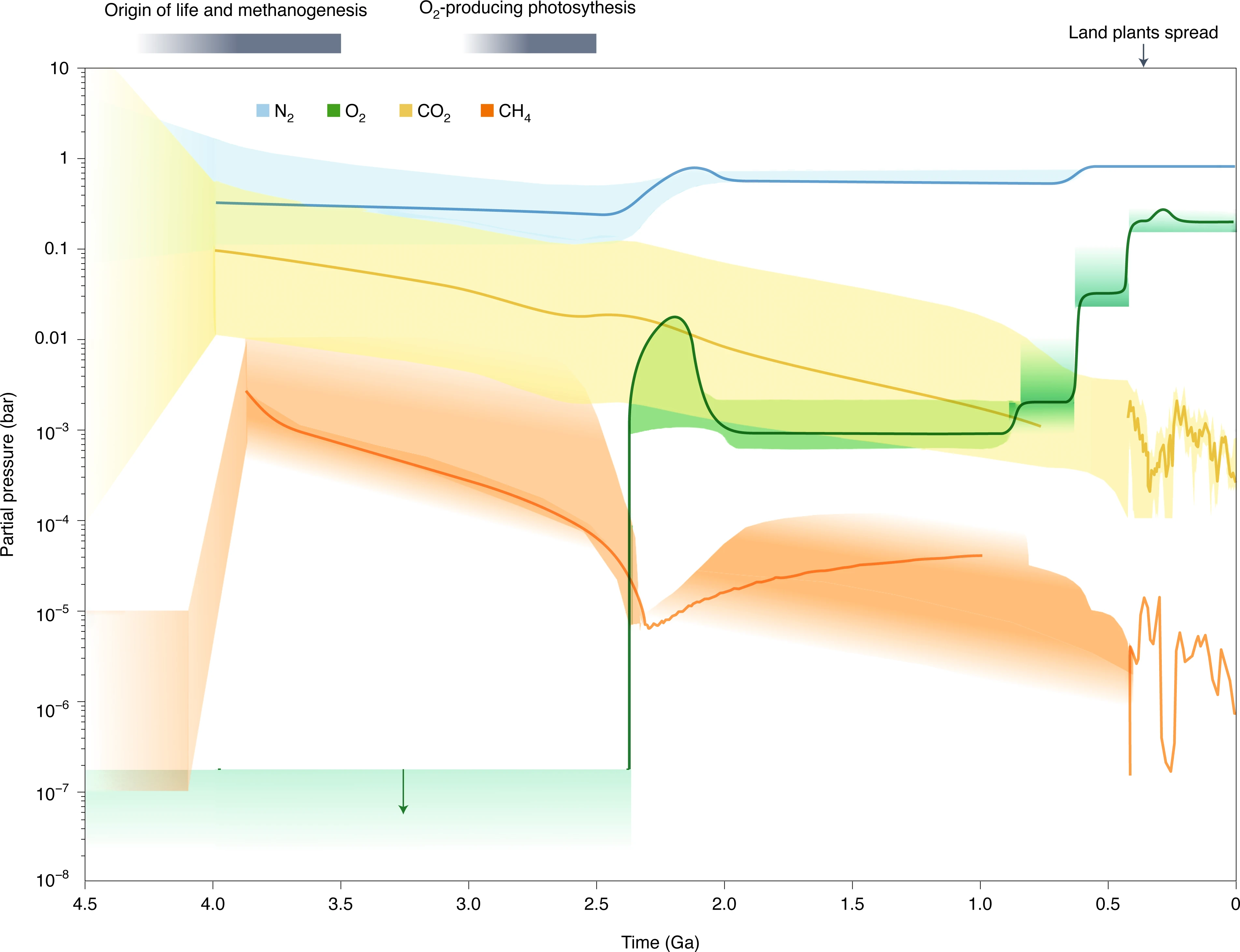}
  \caption{\textbf{Bulk composition of Earth’s atmosphere through time.} Solid lines and shaded regions show the approximate best estimates and approximate uncertainties of N$_2$, O$_2$, CO$_2$ and CH$_4$ partial pressures through time, as determined by both proxies and models. After the emergence of life, atmospheric CH$_4$ was probably abundant throughout Earth’s history, especially before the rise of oxygen. The downward arrow represents an upper limit for O$_2$. Figure adapted from ref.$^{113}$ under a Creative Commons license CC BY 4.0; data for lifeless Hadean atmospheric abundances from refs.$^{55,114}$.}

\end{figure}

The CH$_4$ in Earth’s atmosphere today, and indeed throughout Earth history, is overwhelmingly biogenic, produced by methanogenic microbes via these net reactions:
\begin{equation}
\text{CO$_{2}$ + 4H$_{2}$ $\rightarrow$ CH$_{4}$ + 2H$_{2}$O }
\end{equation}
\begin{equation}
\text{CH$_{3}$COOH $\rightarrow$ CH$_{4}$ + CO$_{2}$}
\end{equation}

Even CH$_4$ that is not directly produced by life is typically a by-product of the degradation of organic matter from previously living organisms (thermogenic CH$_4$). The dominance of biological CH$_4$ sources has led to the suggestion that abundant CH$_4$ could be a sign of life elsewhere, perhaps an indicator of early Earth-like biospheres$^{48,49,50,51,52}$. Recalling life’s universal need for free energy to sustain a far-from-equilibrium state, there is also a strong evolutionary incentive for methanogens to proliferate due to the probable ubiquity of the necessary substrates, CO$_2$+H$_2$, as common outgassing products. Indeed, methanogenesis seems to have emerged easily and early in Earth’s evolution$^{53,54}$. Naturally, the mere presence of methanogenic life does not guarantee detectable atmospheric CH$_4$; low nutrient availability, small H$_2$ degassing fluxes or limited surface habitability could result in false negatives. Nevertheless, ecosystem modelling of diverse chemotrophic and anoxygenic photosynthesizing biospheres provides numerous plausible scenarios for biogenic, CH$_4$-rich atmospheres$^{55,56,57}$.

Methane is a more compelling biosignature when found in combination with CO$_2$ alongside the absence or low abundance of carbon monoxide (CO)$^{58}$. The combination of CH$_4$ and CO$_2$ represents carbon in its most reduced and most oxidized forms, respectively, which is hard to explain without life$^{58}$ (Fig. 4a). Moreover, in terrestrial planet atmospheres, CH$_4$ has a short photochemical lifetime, and thus high abundances require substantial replenishment fluxes$^{50,51}$. Assessment of the biogenicity of CH$_4$ in terrestrial planet atmospheres therefore depends on whether, and in what contexts, it could be produced in large quantities without life. None of the most obvious non-biological replenishment processes can plausibly maintain abundant CH$_4$+CO$_2$ without also producing abundant CO (Fig. 4). For example, magmatic outgassing sufficiently reducing to produce CH$_4$ would also generate large accompanying fluxes of CO (ref.$^{59}$) (Fig. 4b), and for strongly reduced magma compositions, CH$_4$ would probably not outgas to any appreciable extent due to graphite saturation. Similarly, CH$_4$ produced continuously via impacts requires excessive impactor fluxes and is also expected to generate abundant CO (ref.$^{50}$) (Fig. 4c). Very large impacts can produce transient (roughly million year) CH$_4$-rich atmospheres, but such atmospheres are typically H$_2$-dominated and could therefore be identified via low mean molecular weight$^{60}$. Serpentinization and Fischer–Tropsch type reactions are perhaps the most plausible mechanisms for generating a CH$_4$ false positive (Fig. 4d), but in this case the CH$_4$ flux is limited by the supply of fresh crust and by the efficiency of conversion from H$_2$ to CH$_4$. On the modern Earth, serpentinization fluxes of CH$_4$ at mid-ocean ridges and subduction zones are around 2–3 orders of magnitude lower than the global biological flux due to these supply limits$^{61,62,63}$. Although this flux could be higher on planets with greater crustal production and more efficient catalysis of abiotic CO$_2$ reduction, the likelihood of producing abiotic fluxes comparable to Earth’s biological production is still low$^{58}$. Volatile-rich objects (for example Titan-like compositions) may retain sizeable primordial reservoirs of CH$_4$ and CO$_2$ in their icy interiors64 (Fig. 4e), but such CH$_4$ would not persist for more than $\sim$10$^7$–10$^8$ years at habitable zone planet–star separations$^{65}$, and these planets could also be identifiable via their anomalously low densities. Exploration of these conspicuous scenarios (Fig. 4) is a first step towards a contextual framework for CH$_4$ biosignatures, including the identification of observational clues for disentangling biological CH$_4$ from non-biological processes. However, many other geochemical reactions and geophysical processes have been conjectured to contribute to abiotic CH$_4$ production on Earth$^{66}$, and while such fluxes are negligible compared with that of Earth’s biosphere, understanding these proposed mechanisms and extrapolating them to other planetary environments, compositions and tectonic regimes remains an underexplored area of astrobiological research.

\begin{figure}
  \centering
  \includegraphics[width=17cm]{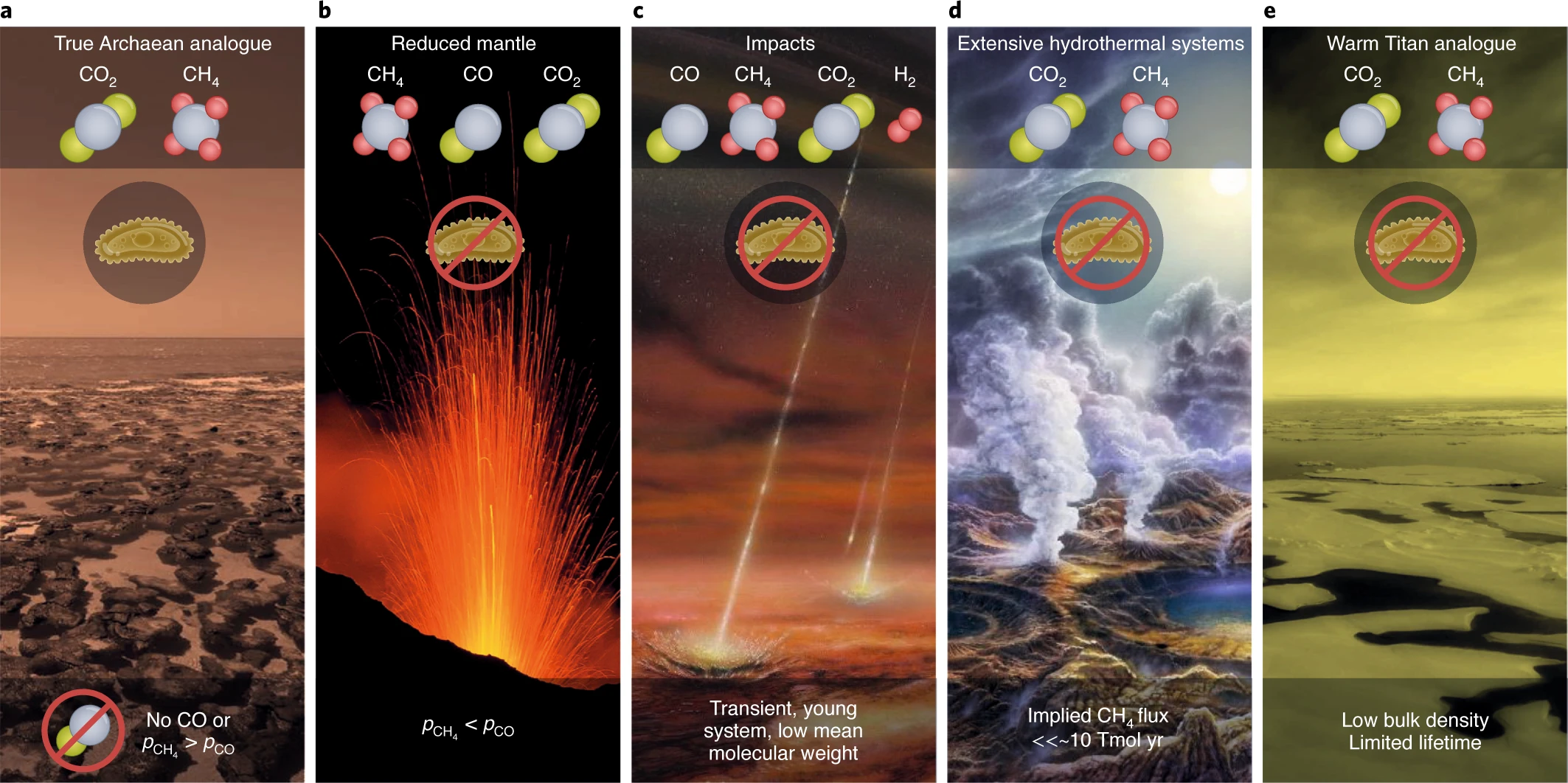}
  \caption{\textbf{Planetary context for CH$_4$ biosignatures and their non-biological false positives.} (a), For productive surface biospheres, the presence of abundant CH$_4$ and CO$_2$ along with the absence or comparatively low abundance of CO suggests a metabolic origin for CH$_4$. (b)–(d), Plausible scenarios for non-biological CH$_4$ production. (b), CH$_4$ from magmatic outgassing requires highly reducing melts that would also produce high CO fluxes$^{59}$. (c), High impact fluxes may generate CH$_4$ transiently, but this CH$_4$ will typically also be accompanied by high CO abundances or H$_2$-dominated atmospheres for transient post-impact atmospheres$^{60}$. (d), Water–rock reactions and Fischer–Tropsch type reactions may produce CH$_4$, but fluxes (and therefore abundances) comparable to those of productive biospheres are improbable$^{58}$; however, the precise threshold will depend on stellar context. (e), Volatile-rich planets could retain large primordial CH$_4$ inventories in subsurface ices, but this is unlikely to sustain high CH$_4$ abundances on geologic lifetimes unless a substantial fraction of the planet is ice$^{65}$. pCH$_4$, partial pressure of CH$_4$; pCO, partial pressure of CO. The bar at the bottom summarizes the contextual clues that could be used to discriminate the five scenarios. Credits (background images): Don Dixon; Donald Hobern; kuhnmi.}
\end{figure}

Note that while abundant CO is generated by many of the non-biological processes in Fig. 4, abundant CO is generally not expected on inhabited planets due to CO being a readily consumed source of free energy for microorganisms$^{67,68}$. This CO antibiosignature argument is complicated by the fact that biospheres dominated by oxygenic photosynthesis could produce high CO fluxes via biomass burning, and that even anoxic biospheres may produce some CO due to incomplete conversion of CO to CH$_4$ (refs.${55,57}$). Further ecosystem modelling is required to determine the range of CO$_2$, CH$_4$ and CO abundances expected for different biospheres and stellar types, but current ecosystem modelling shows that CH$_4$/CO ratios ought to be high on inhabited planets$^{55,56,57}$, perhaps suggesting that the relative abundance of CH$_4$/CO could help distinguish non-biological CH$_4$ from metabolism$^{65}$.

\section{Observational prospects for CH$_4$ biosignatures in the 2020s}

Theoretical understanding of CH$_4$ biosignatures and their false positives may soon lag behind observational capabilities. It is worth noting that CH$_4$ biosignatures are potentially detectable with JWST$^{34,69}$, with simulated retrievals suggesting that biogenic CH$_4$ on TRAPPIST-1e, 1f and 1g may be detectable with approximately 10 transits with NIRSpec prism (Fig. 5). There are even claims of CH$_4$ detections with existing instruments for the highly irradiated terrestrial planet GJ 1132b$^{70}$, although the proposed outgassing mechanism is improbable as it does not account for graphite saturation under reducing conditions$^{65}$, and a featureless spectrum has been suggested by independent analyses of the same Hubble data$^{71,72}$. Methane may also be detectable indirectly via organic haze features in high CH$_4$/CO$_2$ atmospheres$^{52}$.

\begin{figure}
  \centering
  \includegraphics[width=12cm]{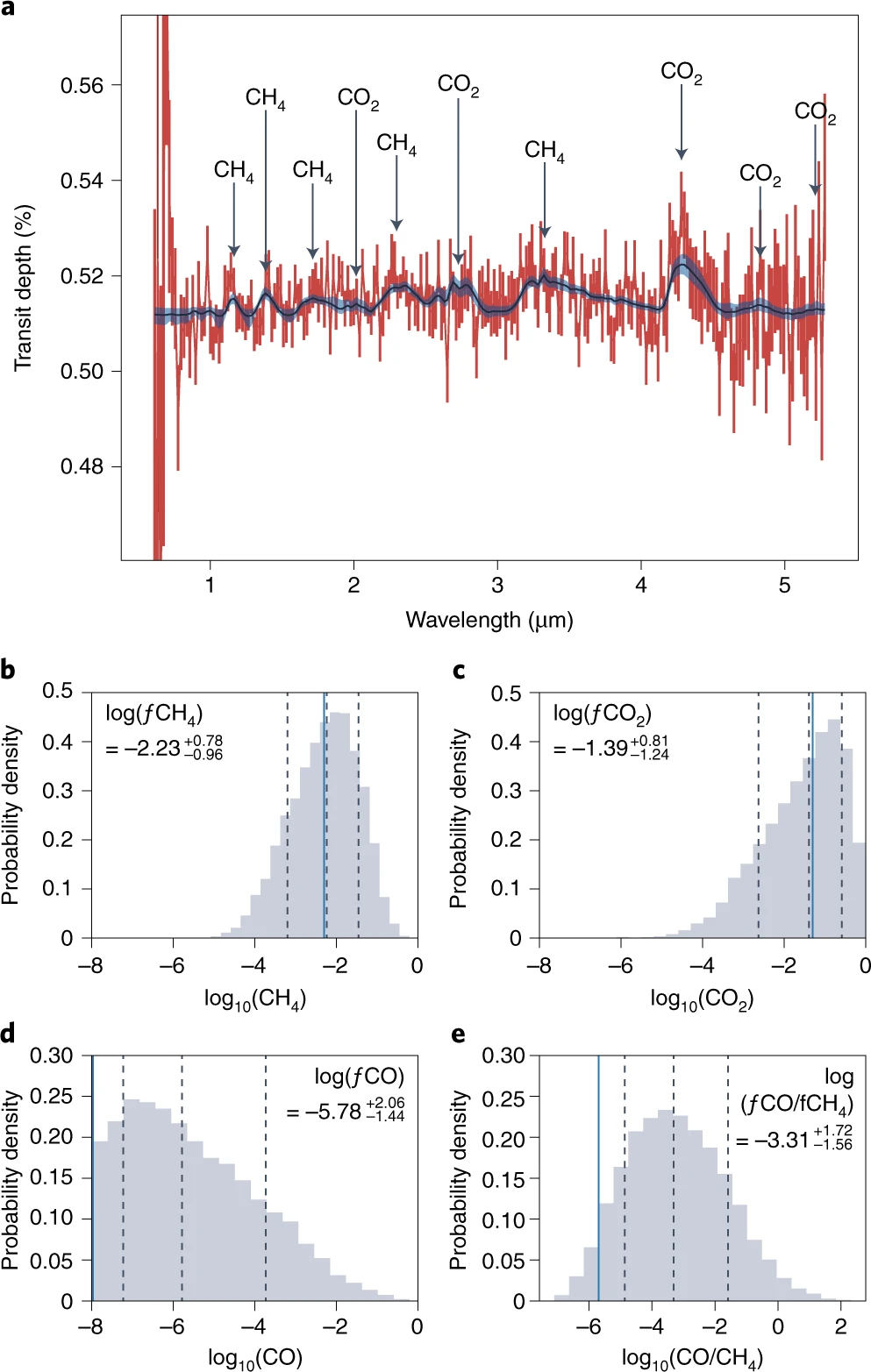}
  \caption{\textbf{Detectability of biogenic CH$_4$ with JWST.} Simulated ten-transit retrieval of an Archaean Earth-like TRAPPIST-1e (Fig. 4a). (a), Simulated noisy transmission spectrum (red) alongside the retrieved spectrum (black) with the 95\% credible interval shaded blue. (b)–(e), Posterior mixing ratios for CH$_4$ (b), CO$_2$ (c), CO (d) and CO/CH$_4$ (e), respectively (the 66\% confidence interval is shown by the dashed black lines) alongside true synthetic abundances (blue lines). The retrieved mixing ratios with 1$\sigma$ uncertainties are included at the tops of the panels, where (f) is the mixing ratio of the indicated species. Biogenic CH$_4$ is detectable, and the CH$_4$ CO$_2$ and CO abundances are sufficiently constrained to disfavour the false positive scenarios in Fig. 4. However, these observations are necessary, but insufficient, to determine biogenicity if an Archaean Earth-like scenario such as Fig. 4a is encountered. Habitability must be confirmed through other means, since atmospheric water vapour and surface properties are inaccessible to JWST transit observations. Figure adapted with permission from ref. $^{34}$, IOP Publishing.}
\end{figure}

What are the prospects for interpreting near-term CH$_4$ detections as biogenic—or, in other words, could life be confirmed on an Archaean Earth analogue such as that depicted in Fig. 4a? Although CO$_2$ detections and crude constraints on CO abundances may be possible$^{34}$, unambiguously ruling out high CO/CH$_4$ ratios consistent with abiotic processes will require many transits (Fig. 5). More fundamentally, JWST will not provide sufficient information on planetary context to fully assess habitability. For example, water vapour is probably inaccessible to transit observations due to cloud condensation$^{73,74}$, and so JWST will be unable to assess the surface conditions on planets with CH$_4$ detections. With that said, the detection of abundant CH$_4$ in terrestrial planet atmospheres with JWST would be tantalizing, especially alongside CO$_2$ and with a high CH$_4$/CO ratio. Such a detection, while not definitive, would undoubtedly motivate the development of future instruments to better characterize planetary context and look for corroborating evidence for a biosphere.

\section{Alternative and agnostic biosignatures}
Whereas the direct products of energy metabolism, such as O$_2$ and CH$_4$, are broadly appreciated as promising context-dependent biosignature gases, there is a growing literature on ‘alternative’ biosignatures. These include other gaseous by-products of metabolism, surface pigments such as the vegetative red edge$^{75,76,77}$, temporal variations in gaseous or surface features indicative of a productive biosphere8, or agnostic biosignatures that are not specific to particular underlying biochemistries but instead invoke generalizable thermodynamic or informational properties of life$^{78,79,80}$.

Virtually all of these potential signs of life are dependent on the whole-planet context: both the evolutionary incentive for their production and the likelihood of non-biological false positives need to be considered. For example, nitrous oxide (N$_2$O) is not a direct product of energy metabolism, but is instead a by-product of incomplete denitrification (the reduction of nitrate to molecular nitrogen). If life elsewhere actively cycles nitrogen, which seems probable given the importance of nitrogen to many organic molecules$^{81}$, then the short photochemical lifetime of N$_2$O makes it a promising biosignature gas$^{9}$. With that said, there are non-biological mechanisms for generating N$_2$O via particle-flux-induced atmospheric chemistry$^{82}$ or even via abiotic iron oxidation in hypersaline cold environments$^{83}$. Thus, any assessment of N$_2$O biogenicity will require constraints on the planetary and stellar context, including an assessment of plausible non-biological fluxes. Similar contextual considerations apply to almost all other gaseous and surface biosignatures that have been proposed. Methyl chloride (CH$_3$Cl) is a by-product of metabolism$^{84}$, but is also potentially produced via metamorphism of organic matter and evaporites$^{85}$ or via volcanic degassing$^{86}$. Phosphine (PH$_3$) may also be a by-product of metabolism in oxidizing atmospheres, whereas its formation in deep, H$_2$-dominated atmospheres is thermodynamically favoured$^{87}$. Seasonal variation in atmospheric O$_2$ or CO$_2$ could be biologically driven, but climate–photochemical variations must be ruled out$^{8}$.

With that said, there are some molecules with formation pathways so thermodynamically improbable that their presence in exoplanet atmospheres in high abundances is extremely unlikely to be attributable to non-biological processes, regardless of the planetary context. Such context-independent biosignatures are more analogous to the kinds of unambiguous in situ Solar System biosignatures discussed above, since they provide direct access to thermodynamic structures and evolutionary properties of life. Technosignatures such as industrial waste products (for example chlorofluorocarbons) fall into this category, as may complex organic molecules such as isoprene$^{88}$ and organo-sulfur compounds such as dimethyl sulfide$^{9,89}$ (while isoprene has never been detected on Titan and is not a predicted photochemical product$^{88}$, the presence of propadiene$^{90}$ suggests that abiotic isoprene synthesis—albeit with trivial fluxes—is not completely implausible.) The obvious disadvantage to these biosignatures is that they are not direct products of energy metabolism but are instead incidental by-products of secondary metabolic processes. As such, they may be specific to Earth life, and the likelihood of large fluxes is challenging to assess. Moreover, greater molecular complexity is typically associated with shorter photochemical lifetimes, leading to poor detectability prospects unless biogenic fluxes are extreme. Isoprene and dimethyl sulfide are also challenging to separate spectroscopically from other hydrocarbons$^{88}$. Although more context-independent biosignatures could provide a shortcut to definitive life detection on exoplanets, the chances of success for any particular biosignature are arguably slim as there is no clear metabolic or evolutionary incentive for biospheres to produce such molecules in detectable quantities.

In fact, the near-term ($\sim$10 yr) detectability prospects for the alternative biosignature gases described above are almost universally bleak$^{87,88,91}$. One possible exception is within extended H$_2$-dominated atmospheres, where the detectability of N$_2$O, CH$_3$Cl, dimethyl sulfide, PH3 and isoprene is enhanced$^{9,92,93}$. The habitability of H$_2$-dominated atmospheres has long been a subject of speculation$^{94,95}$, but these environments remain deeply unfamiliar planetary contexts. Unlike in high-mean-molecular-weight terrestrial atmospheres where silicate weathering feedbacks seemingly stabilize surface habitability against stellar evolution$^{96,97}$, no plausible geochemical feedback mechanism has been proposed for maintaining temperate climates on H$_2$-rich worlds; surface habitability would be serendipitous and probably short-lived without carefully tuned Gaian feedbacks$^{95,98}$. The metabolic incentive to produce the abovementioned alternative biosignature gases in H$_2$-dominated atmospheres is also unclear.

In addition to considering standalone biosignatures, the limitations of N=1 Earth life have motivated explorations of more generalized life metrics that are (ideally) independent of underlying biochemistry, or that make no assumptions about specific metabolisms. Atmospheric chemical disequilibrium has long been debated as a potentially more universal sign of life$^{10,77,99,100}$. The coexistence of O$_2$ and CH$_4$ is certainly a more compelling biosignature than either species alone, due to the short photochemical lifetime of CH$_4$ in O$_2$-rich atmospheres, which would imply high replenishment fluxes for both species that are unlikely to be non-biological. Similarly, as discussed above, the coexistence of CH$_4$ and CO$_2$ is a form of disequilibrium biosignature, although here the argument is focused less on photochemistry than on the redox state of the interior and the low likelihood of large fluxes of both the most oxidized and reduced forms of carbon without biological catalysis. The relationship between atmospheric disequilibrium and life is more nuanced and context-dependent than these examples suggest, however: life depletes free-energy gradients in its environment and sometimes generates disequilibrium as an incidental by-product of incomplete free-energy exploitation, especially when photosynthesis provides a source of free energy external to geochemical processes$^{101}$. In contrast, in other planetary contexts, a large untapped free-energy gradient could be an antibiosignature$^{67}$. A more pragmatic difficulty with photochemical disequilibria, such as the O$_2$ and CH$_4$ pair, is that kinetic instabilities can work against simultaneous detectability$^{99}$, giving rise to possible false negatives or ‘cryptic biospheres’$^{102}$. On the modern Earth, the strongly oxidizing conditions of the modern atmosphere and oceans restrict atmospheric CH$_4$ abundances to a few parts per million. And while O$_2$ and CH$_4$ abundances in the Proterozoic remain uncertain (Fig. 3, but see also Lyons et al.$^{103}$), it is conceivable that photochemical feedbacks and efficient oxidation of biogenic CH$_4$ in the oceans kept both O$_2$ and CH$_4$ abundances below easily detectable thresholds in the mid-Proterozoic$^{102,104}$.

If focusing attention on individual biosignature gases central to energy metabolism (for example O$_2$, CH$_4$) is a top-down approach to biosignatures, then a complementary bottom-up approach would be to systematically characterize all small molecules that could conceivably be produced by life$^{99,105}$. This exhaustive method is more agnostic to underlying biochemistry, and may uncover thermodynamically improbable complex molecules that require less contextual information to be deemed biogenic. On the other hand, this ‘all small molecules’ approach necessitates a huge investment of effort to determine the thermodynamic, kinetic and radiative properties of candidate molecules, many of which lack an obvious evolutionary incentive for planetary-scale biological synthesis—not to mention the work required to characterize their false positives within plausible planetary contexts. Both top-down and bottom-up approaches can yield valuable insights, but given the incomplete contextual understanding and imminent detectability of the biosignature gases central to carbon-based energy metabolism (that is O$_2$ via equation (1) and CH$_4$ via equation (2)), we favour prioritizing an improved understanding of these biosignatures and their false positives.

Alternative ideas for agnostic biosignature metrics include the chemical network properties of atmospheres modulated by life$^{79,106,107}$, where network connectivity or complexity is envisaged as a sign of life independent of specific biochemistry. For all proposed agnostic biosignatures, characterizing the planetary context remains a necessary step for assessing biogenicity; interpretation of atmospheric disequilibrium or network properties will require both an evaluation of how these metrics apply to uninhabited worlds as well as the expected modification by biospheres.

\section{(When) will exoplanet life detection be definitive}
Barring the lucky detection of complex, high-specificity molecules, corroborating context-dependent exoplanet biosignatures will be a gradual process. The possibility of false positives for metabolic products such as O$_2$ and CH$_4$ implies that multiple independent lines of evidence for life may be required for a definitive assessment. For example, the detection of gaseous biosignatures in combination with temporal variations indicative of seasonal metabolic activity alongside reflectance spectra suggestive of surface pigmentation would provide a detailed picture of the planetary context that is challenging to explain without modification by Darwinian processes; non-biological explanations would require an ever-increasing number of ad hoc assumptions.

JWST and the ground-based Extremely Large Telescopes will not have the capabilities to provide such compelling evidence for life. However, they could make the first tentative detections of potential biosignature gases such as CH$_4$ and, more importantly, they will provide an invaluable opportunity to test conceptual frameworks for how terrestrial planets evolve. For example, JWST observations of highly irradiated terrestrial planets will constrain the likelihood of non-biological O$_2$ accumulation on more temperate planets. Crucially, these observations will lay the groundwork for more expansive searches for life with next-generation telescopes$^{1,2,108,109}$, designed with capabilities to constrain the planetary context such that genuine biosignatures and lifeless evolution can be disentangled.

Looking ahead to these future missions, it is worth considering the possible outcomes. If no biosignatures are detected across a large sample of terrestrial planets, and if false negatives can be largely ruled out$^{102}$, then it may be safe to conclude that productive surface biospheres are uncommon. Moreover, if the ostensible habitability of these uninhabited planets can be confirmed$^{30}$, then the most probable explanation for Earth’s uniqueness would be that the origin of life was improbable. The second possible outcome is that multiple independent lines of evidence for life emerge—perhaps even on multiple planets—for which there is no plausible non-biological explanation. In this case, the Copernican Principle will have been reaffirmed beyond reasonable doubt. The third possibility is that the evidence for life on exoplanets remains ambiguous. The limited information gleaned from remote observations and degeneracies in validated models of terrestrial planet evolution could lead to a situation in which the biogenicity of observed planetary environments remains debatable, even with next-generation telescopes such a flagship direct-imaging mission$^{1}$. Even in this scenario, however, new-found quantitative understanding of terrestrial planet evolution could be leveraged to determine what new observations would be required to resolve the ambiguities. There is no shortage of plausible concepts for future telescopes with degeneracy-breaking capabilities, such as the ability to detect trace biosignature gases and their isotopes, or even resolve planetary surfaces$^{110,111,112}$. In short, constraining the abundance of life in the Universe will be possible via astronomical observations, but there is insufficient understanding of how much we need to know about the planetary context to be definitive. The search for life elsewhere could be over soon, or it could take generations, depending on both the resources dedicated to the effort, and the as-yet unknown diversity and complexity of uninhabited worlds that exist in nature.

\section{Conclusions}
The discovery of life on exoplanets, if this occurs, will probably not be a moment in time. It will instead involve years or even decades of debate, with multiple observations and instruments leveraged to resolve ambiguities. No single molecule or surface feature is likely to provide definitive evidence, and multiple lines of evidence may be required to rule out false positives.

Robust life detection will only be possible if testable models of terrestrial planet evolution are developed and validated against a population of lifeless terrestrial planets. This understanding of planetary context will be necessary to confidently rule out non-biological scenarios.

Oxygen is the model biosignature molecule for this contextual approach. Theoretical progress has been made and model predictions for non-biological oxygen accumulation will be testable in the JWST era. Many uncertainties remain, however, especially relating to atmospheric escape physics, planet formation and surface–atmosphere interactions.

For many other potential biosignature gases, understanding of planetary context is lagging behind observability. This is particularly true of CH$_4$, which may be readily detectable with JWST and yet false positive scenarios have not been thoroughly explored. Preliminary investigations of magmatic outgassing, hydrothermal systems and impacts suggest that abundant atmospheric CH$_4$ coexisting with CO$_2$ and limited CO is hard to explain without life, but future study ought to develop coupled geochemical evolution models of a planet’s mantle and crust to self-consistently predict volatile fluxes from both high- and low-temperature systems, such that the contextual clues of non-biological CH$_4$ can be predicted.

Although ambiguities may persist into the space-based direct-imaging era, whole-planet frameworks can be leveraged to determine what contextual observations need to be pursued with future instruments.

\section{Data Availability}
The data outputs from Supplementary Video 1 are available at https://doi.org/10.5281/zenodo.5719456.

\section{Code Availability}
The Python code for our atmosphere evolution model is open source and available at https://doi.org/10.5281/zenodo.4539040.

\section*{Acknowledgments}
J.K.-T. is a NASA Hubble Fellow and was supported by the NASA Sagan Fellowship and through NASA Hubble Fellowship grant number HF2-51437 awarded by the Space Telescope Science Institute, which is operated by the Association of Universities for Research in Astronomy, Inc. for NASA under contract number NAS5-26555. We acknowledge use of the lux supercomputer at UC Santa Cruz, funded by NSF MRI grant number AST 1828315.

\bibliographystyle{unsrt}  
\bibliography{references}

[1] Pathways to Discovery in Astronomy and Astrophysics for the 2020s (The National Academies Press, 2021). (n.d.).

[2] Quanz, S. P. et al. Exoplanet science with a space-based mid-infrared nulling interferometer. Proc. SPIE. https://doi.org/10.1117/12.2312051 (2018).

[3] Baross, J. et al. The Limits of Organic Life in Planetary Systems (National Research Council of the National Academies, 2007).

[4] Schrödinger, E. What is Life? The Physical Aspect of the Living Cell (Cambridge Univ. Press, 1944).

[5] Moberg, C. (2020). Schrödinger’s What is Life?—the 75th anniversary of a book that inspired biology. Angew. Chem. Int. Ed., 132.

[6] Benner, S. A. (2017). Detecting Darwinism from molecules in the Enceladus plumes, Jupiter’s moons, and other planetary water lagoons. Astrobiology, 17.

[7] Hoehler, T. M., Amend, J. P., and Shock, E. L. (2007). A ‘follow the energy’ approach for astrobiology. Astrobiology, 7.

[8] Olson, S. L. (2018). Atmospheric seasonality as an exoplanet biosignature. Astrophys. J. Lett., 858.

[9] Schwieterman, E. W. (2018). Exoplanet biosignatures: a review of remotely detectable signs of life. Astrobiology, 18.

[10] Lovelock, J. E. (1965). A physical basis for life detection experiments. Nature, 207.

[11] Owen, T. in Strategies for the Search for Life in the Universe (ed. Papagiannis, M.D.) 177–185 (Springer, 1980).

[12] Meadows, V. S. (2018). Exoplanet biosignatures: understanding oxygen as a biosignature in the context of its environment. Astrobiology, 18.

[13] Gao, P., Hu, R., Robinson, T. D., Li, C., and Yung, Y. L. (2015). Stabilization of CO2 atmospheres on exoplanets around M dwarf stars. Astrophys. J., 806.

[14] Tian, F., France, K., Linsky, J. L., Mauas, P. J., and Vieytes, M. C. (2014). High stellar FUV/NUV ratio and oxygen contents in the atmospheres of potentially habitable planets. Earth Planet. Sci. Lett., 385.

[15] Harman, C., Schwieterman, E., Schottelkotte, J. C., and Kasting, J. (2015). Abiotic O2 levels on planets around F, G, K, and M stars: possible false positives for life? Astrophys. J., 812.

[16] Domagal-Goldman, S. D., Segura, A., Claire, M. W., Robinson, T. D., and Meadows, V. S. (2014). Abiotic ozone and oxygen in atmospheres similar to prebiotic Earth. Astrophys. J., 792.

[17] Harman, C. (2018). Abiotic O2 levels on planets around F, G, K, and M stars: effects of lightning-produced catalysts in eliminating oxygen false positives. Astrophys. J., 866.

[18] Hu, R., Peterson, L., and Wolf, E. T. (2020). O2-and CO-rich atmospheres for potentially habitable environments on TRAPPIST-1 planets. Astrophys. J., 888.

[19] Grenfell, J. L. (2018). Limitation of atmospheric composition by combustion–explosion in exoplanetary atmospheres. Astrophys. J., 861.

[20] Luger, R., and Barnes, R. (2015). Extreme water loss and abiotic O2 buildup on planets throughout the habitable zones of M dwarfs. Astrobiology, 15.

[21] Schaefer, L., Wordsworth, R. D., Berta-Thompson, Z., and Sasselov, D. (2016). Predictions of the atmospheric composition of GJ 1132b. Astrophys. J., 829.

[22] Wordsworth, R., Schaefer, L., and Fischer, R. (2018). Redox evolution via gravitational differentiation on low-mass planets: implications for abiotic oxygen, water loss, and habitability. Astron. J., 155.

[23] Barth, P. (2021). Magma ocean evolution of the TRAPPIST-1 planets. Astrobiology, 21.

[24] Wordsworth, R., and Pierrehumbert, R. (2014). Abiotic oxygen-dominated atmospheres on terrestrial habitable zone planets. Astrophys. J. Lett., 785.

[25] Kleinböhl, A., Willacy, K., Friedson, A. J., Chen, P., and Swain, M. R. (2018). Buildup of abiotic oxygen and ozone in moist atmospheres of temperate terrestrial exoplanets and its impact on the spectral fingerprint in transit observations. Astrophys. J., 862.

[26] Krissansen-Totton, J., Fortney, J. J., Nimmo, F. and Wogan, N. Oxygen false positives on habitable zone planets around sun-like stars. AGU Adv.2, e2020AV000294 (2021). 

[27] Noack, L. (2016). Water-rich planets: how habitable is a water layer deeper than on Earth? Icarus, 277.

[28] Kite, E. S., and Ford, E. B. (2018). Habitability of exoplanet waterworlds. Astrophys. J., 864.

[29] Cowan, N. B., and Abbot, D. S. (2014). Water cycling between ocean and mantle: super-Earths need not be waterworlds. Astrophys. J., 781.

[30] Lustig-Yaeger, J. (2018). Detecting ocean glint on exoplanets using multiphase mapping. Astron. J., 156.

[31] Fujii, Y. (2010). Colors of a second Earth: estimating the fractional areas of ocean, land, and vegetation of Earth-like exoplanets. Astrophys. J., 715.

[32] Cowan, N. B. (2009). Alien maps of an ocean-bearing world. Astrophys. J., 700.

[33] Fauchez, T. J. (2020). Sensitive probing of exoplanetary oxygen via mid-infrared collisional absorption. Nat. Astron., 4.

[34] Krissansen-Totton, J., Garland, R., Irwin, P., and Catling, D. C. (2018). Detectability of biosignatures in anoxic atmospheres with the James Webb Space Telescope: a TRAPPIST-1e case study. Astron. J., 156.

[35] Lustig-Yaeger, J., Meadows, V. S., and Lincowski, A. P. (2019). The detectability and characterization of the TRAPPIST-1 exoplanet atmospheres with JWST. Astron. J., 158.

[36] Wunderlich, F. (2019). Detectability of atmospheric features of Earth-like planets in the habitable zone around M dwarfs. Astron. Astrophys., 624.

[37] Rodler, F., and López-Morales, M. (2014). Feasibility studies for the detection of O2 in an Earth-like exoplanet. Astrophys. J., 781.

[38] Snellen, I., Kok, R., Poole, R., Brogi, M., and Birkby, J. (2013). Finding extraterrestrial life using ground-based high-dispersion spectroscopy. Astrophys. J., 764.

[39] Leung, M., Meadows, V. S., and Lustig-Yaeger, J. (2020). High-resolution spectral discriminants of ocean loss for M-dwarf terrestrial exoplanets. Astron. J., 160.

[40] Lincowski, A. P. (2018). Evolved climates and observational discriminants for the TRAPPIST-1 planetary system. Astrophys. J., 867.

[41] Gillon, M. (2017). Seven temperate terrestrial planets around the nearby ultracool dwarf star TRAPPIST-1. Nature, 542.

[42] Gillon, M. et al. The TRAPPIST-1 JWST Community Initiative. Preprint at https://arxiv.org/abs/2002.04798 (2020). 

[43] Donahue, T., Hoffman, J., Hodges, R., and Watson, A. (1982). Venus was wet: a measurement of the ratio of deuterium to hydrogen. Science, 216.

[44] Kiefer, W. S. et al. Venus, Earth’s divergent twin?: Testing evolutionary models for Venus with the DAVINCI+ mission. In European Planetary Science Congress 2020 EPSC2020-2534 (EPS, 2020).

[45] Smrekar, S. et al. VERITAS (Venus Emissivity, Radio Science, InSAR, Topography And Spectroscopy): A proposed discovery mission. In European Planetary Science Congress 2020 EPSC2020-447 (EPS, 2020).

[46] Widemann, T., Titov, D., Wilson, C., and Ghail, R. EnVision: Europe’s proposed mission to Venus. In 43rd COSPAR Scientific Assembly (COSPAR, 2021).

[47] Lehmer, O. R., Catling, D. C., Parenteau, M. N., and Hoehler, T. M. (2018). The productivity of oxygenic photosynthesis around cool, M dwarf stars. Astrophys. J., 859.

[48] Schindler, T. L., and Kasting, J. F. (2000). Synthetic spectra of simulated terrestrial atmospheres containing possible biomarker gases. Icarus, 145.

[49] Des Marais, D. J. (2002). Remote sensing of planetary properties and biosignatures on extrasolar terrestrial planets. Astrobiology, 2.

[50] Kasting, J. F. (2005). Methane and climate during the Precambrian era. Precamb. Res., 137.

[51] Guzmán-Marmolejo, A., Segura, A., and Escobar-Briones, E. (2013). Abiotic production of methane in terrestrial planets. Astrobiology, 13.

[52] Arney, G., Domagal-Goldman, S. D., and Meadows, V. S. (2018). Organic haze as a biosignature in anoxic Earth-like atmospheres. Astrobiology, 18.

[53] Weiss, M. C. (2016). The physiology and habitat of the last universal common ancestor. Nat. Microbiol., 1.

[54] Wolfe, J. M., and Fournier, G. P. (2018). Horizontal gene transfer constrains the timing of methanogen evolution. Nat. Ecol. Evol., 2.

[55] Sauterey, B. (2020). Co-evolution of primitive methane-cycling ecosystems and early Earth’s atmosphere and climate. Nat. Commun., 11.

[56] Kharecha, P., Kasting, J., and Siefert, J. (2005). A coupled atmosphere–ecosystem model of the early Archean Earth. Geobiology, 3.

[57] Schwieterman, E. W. (2019). Rethinking CO antibiosignatures in the search for life beyond the Solar System. Astrophys. J., 874.

[58] Krissansen-Totton, J., Olson, S., and Catling, D. C. (2018). Disequilibrium biosignatures over Earth history and implications for detecting exoplanet life. Sci. Adv., 4.

[59] Wogan, N., Krissansen-Totton, J., and Catling, D. C. (2020). Abundant atmospheric methane from volcanism on terrestrial planets is unlikely and strengthens the case for methane as a biosignature. Planet. Sci. J., 1.

[60] Zahnle, K. J., Lupu, R., Catling, D. C., and Wogan, N. (2020). Creation and evolution of impact-generated reduced atmospheres of early Earth. Planet. Sci. J., 1.

[61] Keir, R. (2010). A note on the fluxes of abiogenic methane and hydrogen from mid-ocean ridges. Geophys. Res. Lett., 37.

[62] Cannat, M., Fontaine, F. and Escartin, J. in Diversity of Hydrothermal Systems on Slow Spreading Ocean Ridges 241–264 (American Geophysical Union, 2010).

[63] Vitale Brovarone, A. (2017). Massive production of abiotic methane during subduction evidenced in metamorphosed ophicarbonates from the Italian Alps. Nat. Commun., 8.

[64] Tobie, G., Gautier, D., and Hersant, F. (2012). Titan’s bulk composition constrained by Cassini-Huygens: implication for internal outgassing. Astrophys. J., 752.

[65] Thompson, M. A., Krissansen-Totton, J., Wogan, N. and Fortney, J. J. (2022). The case and context for atmospheric methane as an exoplanet biosignature. Proc. Natl Acad. Sci. USA (in the press).

[66] Etiope, G., and Sherwood Lollar, B. (2013). Abiotic methane on Earth. Rev. Geophys., 51.

[67] Sholes, S. F., Krissansen-Totton, J., and Catling, D. C. (2019). A maximum subsurface biomass on Mars from untapped free energy: CO and H2 as potential antibiosignatures. Astrobiology, 19.

[68] Zahnle, K., Freedman, R. S., and Catling, D. C. (2011). Is there methane on Mars? Icarus, 212.

[69] Mikal-Evans, T. (2022). Detecting the proposed CH4–CO2 biosignature pair with the James Webb Space Telescope: TRAPPIST-1e and the effect of cloud/haze. Mon. Not. R. Astron. Soc., 510.

[70] Swain, M. R. (2021). Detection of an atmosphere on a rocky exoplanet. Astron. J., 161.

[71] Mugnai, L. V. (2021). ARES.* V. No evidence for molecular absorption in the HST WFC3 spectrum of GJ 1132 b. Astron. J., 161.

[72] Libby-Roberts, J. E. et al. The featureless HST/WFC3 transmission spectrum of the rocky exoplanet GJ 1132b: no evidence for a cloud-free primordial atmosphere and constraints on starspot contamination. Preprint at https://arxiv.org/abs/2105.10487 (2021).

[73] Komacek, T. D., Fauchez, T. J., Wolf, E. T., and Abbot, D. S. (2020). Clouds will likely prevent the detection of water vapor in JWST transmission spectra of terrestrial exoplanets. Astrophys. J. Lett., 888.

[74] Fauchez, T. J. (2019). Impact of clouds and hazes on the simulated JWST transmission spectra of habitable zone planets in the TRAPPIST-1 system. Astrophys. J., 887.

[75] Kiang, N. Y., Siefert, J., and Blankenship, R. E. (2007). Spectral signatures of photosynthesis. I. Review of Earth organisms. Astrobiology, 7.

[76] Seager, S., Turner, E. L., Schafer, J., and Ford, E. B. (2005). Vegetation’s red edge: a possible spectroscopic biosignature of extraterrestrial plants. Astrobiology, 5.

[77] Sagan, C., Thompson, W. R., Carlson, R., Gurnett, D., and Hord, C. (1993). A search for life on Earth from the Galileo spacecraft. Nature, 365.

[78] Marshall, S. M. (2021). Identifying molecules as biosignatures with assembly theory and mass spectrometry. Nat. Commun., 12.

[79] Walker, S. I. (2018). Exoplanet biosignatures: future directions. Astrobiology, 18.

[80] Bartlett, S. et al. Assessing planetary complexity and potential agnostic biosignatures using epsilon machines. Nat. Astron. https://doi.org/10.1038/s41550-021-01559-x (2022). 

[81] Capone, D. G., Popa, R., Flood, B., and Nealson, K. H. (2006). Follow the nitrogen. Science, 312.

[82] Airapetian, V., Glocer, A., Gronoff, G., Hebrard, E., and Danchi, W. (2016). Prebiotic chemistry and atmospheric warming of early Earth by an active young Sun. Nat. Geosci., 9.

[83] Samarkin, V. A. (2010). Abiotic nitrous oxide emission from the hypersaline Don Juan Pond in Antarctica. Nat. Geosci., 3.

[84] Keene, W. C. (1999). Composite global emissions of reactive chlorine from anthropogenic and natural sources: reactive chlorine emissions inventory. J. Geophys. Res. Atmos., 104.

[85] Aarnes, I., Fristad, K., Planke, S. and Svensen, H. The impact of host‐rock composition on devolatilization of sedimentary rocks during contact metamorphism around mafic sheet intrusions. Geochem. Geophys. Geosyst. https://doi.org/10.1029/2011GC003636 (2011).

[86] Frische, M., Garofalo, K., Hansteen, T. H. and Borchers, R. Fluxes and origin of halogenated organic trace gases from Momotombo volcano (Nicaragua). Geochem. Geophys. Geosyst. https://doi.org/10.1029/2005GC001162 (2006). 

[87] Sousa-Silva, C. (2020). Phosphine as a biosignature gas in exoplanet atmospheres. Astrobiology, 20.

[88] Zhan, Z. et al. Assessment of isoprene as a possible biosignature gas in exoplanets with anoxic atmospheres. Astrobiology https://doi.org/10.1089/ast.2019.2146 (2021).

[89] Pilcher, C. B. (2003). Biosignatures of early Earths. Astrobiology, 3.

[90] Lombardo, N. A. (2019). Detection of propadiene on Titan. Astrophys. J. Lett., 881.

[91] Gialluca, M. T., Robinson, T. D., Rugheimer, S., and Wunderlich, F. (2021). Characterizing atmospheres of transiting Earth-like exoplanets orbiting M dwarfs with James Webb Space Telescope. Publ. Astron. Soc. Pac., 133.

[92] Seager, S., Bains, W., and Hu, R. (2013). Biosignature gases in H2-dominated atmospheres on rocky exoplanets. Astrophys. J., 777.

[93] Wunderlich, F. (2021). Detectability of biosignatures on LHS 1140 b. Astron. Astrophys., 647.

[94] Stevenson, D. J. (1999). Life-sustaining planets in interstellar space? Nature, 400.

[95] Pierrehumbert, R., and Gaidos, E. (2011). Hydrogen greenhouse planets beyond the habitable zone. Astrophys. J. Lett., 734.

[96] Walker, J. C., Hays, P., and Kasting, J. F. (1981). A negative feedback mechanism for the long‐term stabilization of Earth’s surface temperature. J. Geophys. Res. Oceans, 86.

[97] Lehmer, O. R., Catling, D. C., and Krissansen-Totton, J. (2020). Carbonate-silicate cycle predictions of Earth-like planetary climates and testing the habitable zone concept. Nat. Commun., 11.

[98] Abbot, D. S. (2015). A proposal for climate stability on H2-greenhouse planets. Astrophys. J. Lett., 815.

[99] Seager, S., and Bains, W. (2015). The search for signs of life on exoplanets at the interface of chemistry and planetary science. Sci. Adv., 1.

[100] Krissansen-Totton, J., Bergsman, D. S., and Catling, D. C. (2016). On detecting biospheres from chemical thermodynamic disequilibrium in planetary atmospheres. Astrobiology, 16.

[101] Wogan, N. F., and Catling, D. C. (2020). When is chemical disequilibrium in Earth-like planetary atmospheres a biosignature versus an anti-biosignature? Disequilibria from dead to living worlds. Astrophys. J., 892.

[102] Reinhard, C. T., Olson, S. L., Schwieterman, E. W., and Lyons, T. W. (2017). False negatives for remote life detection on ocean-bearing planets: lessons from the early Earth. Astrobiology, 17.

[103] Lyons, T. W., Diamond, C. W., Planavsky, N. J., Reinhard, C. T., and Li, C. (2021). Oxygenation, life, and the planetary system during Earth’s middle history: an overview. Astrobiology, 21.

[104] Olson, S. L., Reinhard, C. T., and Lyons, T. W. (2016). Limited role for methane in the mid-Proterozoic greenhouse. Proc. Natl Acad. Sci. USA, 113.

[105] Seager, S., Bains, W., and Petkowski, J. (2016). Toward a list of molecules as potential biosignature gases for the search for life on exoplanets and applications to terrestrial biochemistry. Astrobiology, 16.

[106] Estrada, E. (2012). Returnability as a criterion of disequilibrium in atmospheric reactions networks. J. Math. Chem., 50.

[107] Fisher, T., Kim, H., Millsaps, C., Line, M. and Walker, S. I. Inferring exoplanet disequilibria with multivariate information in atmospheric reaction networks. Preprint at https://arxiv.org/abs/2104.09776 (2021).

[108] Fischer, D. et al. The LUVOIR Mission Concept Study Final Report (NASA, 2019).

[109] Gaudi, B. S. et al. The Habitable Exoplanet Observatory (HabEx) mission concept study final report. Preprint at https://arxiv.org/abs/2001.06683 (2020).

[110] Labeyrie, A. (2021). Lunar optical interferometry and hypertelescope for direct imaging at high resolution. Phil. Trans. R. Soc. A, 379.

[111] Turyshev, S. G. et al. Recognizing the value of the solar gravitational lens for direct multipixel imaging and spectroscopy of an exoplanet. Preprint at https://arxiv.org/abs/1803.04319 (2018).

[112] Labeyrie, A. (1999). Snapshots of alien worlds–the future of interferometry. Science, 285.

[113] Catling, D. C., and Zahnle, K. J. (2020). The Archean atmosphere. Sci. Adv., 6.

[114] Kadoya, S., Krissansen‐Totton, J., and Catling, D. C. (2020). Probable cold and alkaline surface environment of the Hadean Earth caused by impact ejecta weathering. Geochem. Geophys. Geosyst., 21.

\end{document}